\documentclass{aa}
\usepackage{txfonts}
\usepackage{graphicx}

 \newcommand{\be}{\begin{equation}}
 \newcommand{\ee}{\end{equation}}
 \newcommand{\bdm}{\begin{displaymath}}
 \newcommand{\edm}{\end{displaymath}}
 \renewcommand{\vec}[1]{\mbox{\boldmath$#1$}}

\begin{document}

   \title{Subsonic propellers in a strong wind as anomalous X-ray pulsars}

   \author{N.R.\,Ikhsanov\inst{1,2,3} and C.-S.\,Choi\inst{1}}

  \offprints{N.R.~Ikhsanov \\ \email{ikhsanov@kao.re.kr}}

   \institute{Korea Astronomy Observatory, 61-1 Hwaam, Yusong,
              Taejon 305-348, Korea
              \and
              Central Astronomical Observatory of
              the Russian Academy of Sciences,
              Pulkovo 65--1, 196140 Saint-Petersburg, Russia
              \and
              Isaac Newton Institute of Chile, St.Petersburg Branch, Russia}

   \date{Received ; accepted }

\authorrunning{N.R.\,Ikhsanov and C.-S.\,Choi}
\titlerunning{Subsonic propellers as anomalous X-ray pulsars}

    \abstract{The appearance of subsonic propellers situated in a strong wind
    is discussed. We show that it is similar to the appearance of anomalous
    X-ray pulsars (AXPs) provided the mass and the magnetic moment
    of neutron stars are $M\simeq 1.4M_{\sun}$, and
    $\mu\simeq 2\times 10^{30}\,{\rm G\,cm^3}$, respectively, and the strength
    of the wind is $\dot{M}_{\rm c} \simeq 3\times 10^{17}\,{\rm g\,s^{-1}}$.
    Under these conditions, the spin periods of subsonic propellers are limited
    within the range of 5--15\,s, and the expected spin-down rates are close
    to $\dot{P}\simeq 7\times 10^{-11}\,{\rm s\,s^{-1}}$. The mass
    accretion rate onto the stellar surface is limited to the rate of
    plasma penetration into its magnetosphere at the boundary. As this process is
    governed by the reconnection of the field lines, the accretion rate
    onto the stellar surface constitutes 1--2\% of $\dot{M}_{\rm
    c}$. In this case the X-ray luminosity of the
    objects under consideration can be evaluated as
    $L_{\rm X} \sim 4\times 10^{35}\,{\rm erg\,s^{-1}}$. The model predicts the
    existence of at least two
    spatially separated sources of the X-ray emission: hot spots at the stellar
    surface, and the hot atmosphere surrounding the magnetosphere of the
    star. The ages of the subsonic propellers under the conditions of interest
    are limited to $\la 10^5$\,yr.  }

   \maketitle


   \section{Introduction}

Anomalous X-ray pulsars (AXPs) are currently recognized as a
separate class of single magnetized neutron stars, with common
characteristics different from those of the normal pulsars. The
theoretical interpretation of these objects is still a point of
discussion. A successful theoretical model of AXPs should be able
to explain the following basic properties of these objects (see
Mereghetti et~al. \cite{mcis02} for a
comprehensive review): \\
(1)~the clustering of spin periods in the range 5--12\,s;\\
(2)~the spin-down rates of $10^{-10}-10^{-12}\,{\rm
s\,s^{-1}}$;\\
(3)~the soft X-ray spectra, which are clearly different from
those of the accretion-powered pulsars; \\
(4)~the X-ray luminosity at a level of $\sim
10^{35}-10^{36}\,{\rm erg\,s^{-1}}$;\\
(5)~relatively young ages of these objects, derived from the
association of at least two of AXPs with supernova shell remnants
(SNRs).

A key question in the modeling of AXPs is the state of the neutron
star. Actually, there are only 4 possible answers: {\it ejector},
{\it supersonic propeller}, {\it subsonic propeller}, and {\it
accretor} (see e.g. Ikhsanov \cite{i01a}). Three of these
possibilities have already been discussed in the literature. The
state of ejector (= spin-powered pulsar) was analyzed within the
magnetar model. The observed spin-down rates, the soft spectra and
the young ages of AXPs can be interpreted within this model
provided the strength of the dipole magnetic field of the star is
$B \ga 10^{15}$\,G (see e.g. Thompson \& Duncan \cite{td96} and
references therein). At the same time, the clustering of the
periods can hardly be explained in terms of magnetars unless some
deviations from the canonical scheme of the evolution of neutron
stars are invoked in the model (for discussion see Psaltis \&
Miller \cite{pm02} and Colpi et~al. \cite{cgp00}).

The excess of the observed luminosity over the spin-down power
allows us to reject the possibility that AXPs are {\it supersonic
propellers}, i.e. spinning-down neutron stars for which the
condition $R_{\rm cor} \la R_{\rm m} < \min\{R_{\alpha}, R_{\rm
lc}\}$ is satisfied. Here $R_{\rm cor}=(GM_{\rm
ns}/\omega^2)^{1/3}$, $R_{\alpha}=2GM_{\rm ns}/V_{\rm rel}^2$ and
$R_{\rm m}$  are the corotation, accretion, and magnetospheric
radii of the neutron star, respectively. $R_{\rm lc}=c/\omega$ is
the radius of the light cylinder. $\omega=2\pi/P_{\rm s}$ and
$M_{\rm ns}$ denote the angular velocity and the mass of the star,
and $V_{\rm rel}$ is the relative velocity between the star and
the surrounding medium. $G$ and $c$ are the gravitational constant
and the speed of light. On this basis, the corresponding state is
presently discussed (e.g. Marsden et~al. \cite{mlrh01}) only in
the context of a previous spin-down evolution of AXPs to their
present state.

The modeling of AXPs as neutron stars accreting material from a
disk has been reported by many authors (see e.g. Chatterjee,
Hernquist and Narayan \cite{chn00}, and Mereghetti et~al.
\cite{mcis02} for a detailed description). Within this approach,
the observed luminosity is interpreted in terms of the accretion
power. Therefore, the `power deficit' problem mentioned above
turns out to be naturally avoided. However, to meet other
criteria, such as the relatively high spin-down rates, the soft
X-ray spectrum and the clustering of periods in a relatively
narrow range, a rather specific scheme of neutron star evolution
has to be invoked (Eksi \& Alpar \cite{ea03}).

In the light of this situation, a consideration of the last
possibility, namely, that the state of neutron stars in AXPs is
{\it subsonic propeller} appears quite reasonable. The
corresponding analysis is the subject of this paper. As shown
below, the modeling of AXPs within the subsonic propeller approach
appears effective in several important aspects. In particular, it
provides us with an explanation of the observed clustering of
periods (Sect.\,2) and spin-down rates (Sect.\,3). It also
predicts the existence of two spatially separated sources of X-ray
emission (Sect.\,4) with the total luminosity of order of the
observed luminosity of AXPs. The ages of supersonic propellers are
evaluated in Sect.\,5. The proposed scenario is discussed in
Sect.\,6 and the basic conclusions are summarized in Sect.\,7.

   \section{Period clustering}\label{pc}

As recently shown by Ikhsanov (\cite{i01b}), the spin period of a
neutron star in the state of {\it subsonic propeller} satisfies
the following condition: $P_{\rm cd} \la P_{\rm s} \la P_{\rm
br}$, where
 \be\label{pcd}
P_{\rm cd} \simeq 3.2\ \mu_{30}^{6/7}\ m^{-5/7}\
\dot{M}_{17}^{-3/7}\ {\rm s},
   \ee
is the period at which the star switches its state from {\it
supersonic} to {\it subsonic} propeller, and
   \be\label{pbr}
P_{\rm br} \simeq\ 16\ \mu_{30}^{16/21}\ \dot{M}_{17}^{-5/7}\
m^{-4/21}\ {\rm s},
      \ee
is the period at which the state of the star switches to {\it
accretor}. Here $m$ and $\mu_{30}$ are the mass and the dipole
magnetic moment of the star expressed in units of $M_{\sun}$, and
$10^{30}\,{\rm G\,cm^3}$, respectively. $\dot{M}_{17}$ is the
strength of the wind, $M_{\rm c}$, expressed in units of
$10^{17}\,{\rm g\,s^{-1}}$. The meaning of this parameter is the
mass of the surrounding material interacting with the neutron
star, moving through the medium with the velocity $V_{\rm rel}$,
in a time unit: $\dot{M}_{\rm c} = \pi R_{\alpha}^2 \rho_{\infty}
V_{\rm rel}$. Here $\rho_{\infty}$ is the density of material
surrounding the star at the distance $R_{\alpha}$.

Combining Eqs.~(\ref{pcd}) and (\ref{pbr}) we find that the
clustering of periods of AXPs can be explained in terms of the
subsonic propeller approach provided these objects are neutron
stars of mass $M\sim 1.4\,M_{\sun}$ and magnetic moment $\mu \sim
2\times 10^{30}\,{\rm G\,cm^3}$, situated in a wind of strength
$\dot{M}_{\rm c} \sim 3\times 10^{17}\,{\rm g\,s^{-1}}$.

  \section{Spin-down rate}

As shown by Davies \& Pringle (\cite{dp81}), the spin-down power
of stars in the state of subsonic propeller can be evaluated as
   \be\label{lssp}
L_{\rm ssp} = 8 \times 10^{33}\ \mu_{30}^2\ m^{-1}\ P_{5}^{-3}\
{\rm egr\,s^{-1}},
   \ee
where $P_{5}=P_{\rm s}/5$\,s. This means that subsonic propellers
are expected to spin down at a rate
 \be\label{dotpsubp}
 \dot{P} = \frac{P_{\rm s}^3 L_{\rm ssp}}{4 \pi^2 I} \sim 2.5
 \times 10^{-11}\  \mu_{30}^2\ m^{-1}\ I_{45}^{-1}\ {\rm
 s\,s^{-1}},
 \ee
where $I_{45}$ is the moment of inertia of the neutron star
expressed in units of $10^{45}\,{\rm g\,cm^2}$.

Putting the above derived parameters ($M\sim 1.4\,M_{\sun}$ and
$\mu \sim 2\times 10^{30}\,{\rm G\,cm^3}$) into
Eq.~(\ref{dotpsubp}), one finds that the spin-down rate of AXPs
within the subsonic propeller approach is expected to be of order
$7 \times 10^{-11}\,{\rm s\,s^{-1}}$. This is in a good agreement
with the typically observed values.

  \section{Accretion power}

According to the picture presented by Davies \& Pringle
(\cite{dp81}), the magnetosphere of a neutron star in the state of
subsonic propeller is surrounded by an adiabatic ($p\propto
R^{-5/2}$) spherically symmetrical plasma envelope. The energy
input to the envelope due to the propeller action by the star
dominates the energy losses from the envelope as long as $P_{\rm
s} < P_{\rm br}$. Therefore, the temperature of the envelope
plasma remains of order of the free-fall temperature, $T_{\rm
pl}\simeq T_{\rm ff}=GM_{\rm ns}m_{\rm p}/kR$ and,
correspondingly, the sound speed is of order of the free-fall
velocity, $V_{\rm s} \simeq V_{\rm ff}=\sqrt{2GM_{\rm ns}/R}$
(here $m_{\rm p}$, and $k$ are the proton mass and the Boltzmann
constant). Under these conditions the height of the homogeneous
atmosphere throughout the envelope is comparable to the radius,
$R$, and the envelope is extended from the magnetospheric boundary
up to the accretion radius of the star.

\begin{figure}
   \centering
\includegraphics[width=8cm]{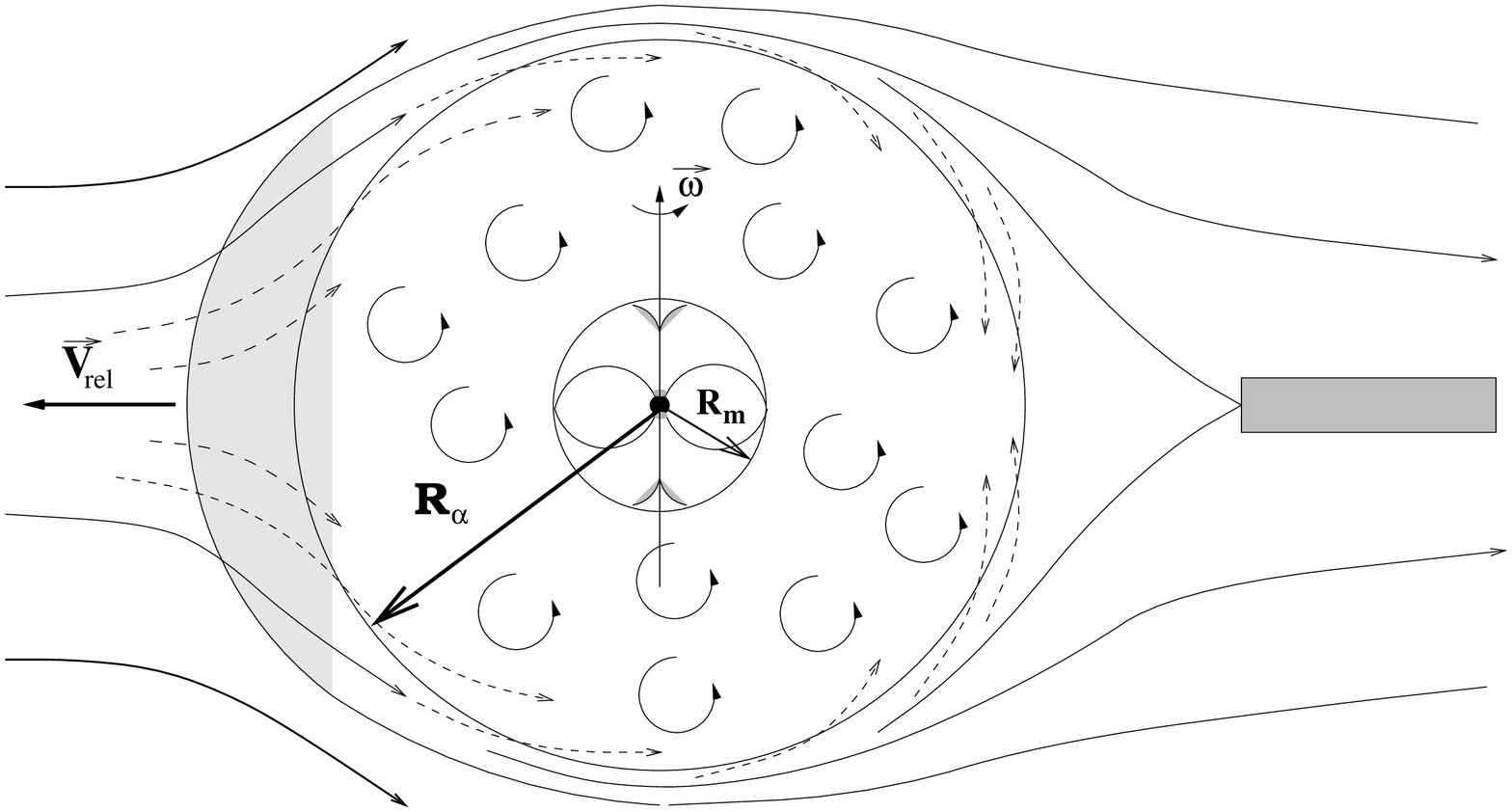}
      \caption{Schematic description of the subsonic propeller. A
 magnetized neutron star rotates with angular velocity
 \vec{\omega}. The magnetosphere of the star is surrounded by a hot
 turbulent envelope. As the star moves through the medium with
 relative velocity ${\bf V}_{\rm rel}$, the gas overflows the outer
 edge of the envelope compressing the envelope plasma. The overflow
 rate is equal to the wind strength, $\dot{M}_{\rm c}$. The rate of
 plasma motion through the envelope is equal to the rate of plasma
 penetration into the magnetosphere of the star at the inner
 boundary of the envelope, $\dot{M}_{\rm rec}$. The material
 penetrating into the magnetic field of the star flows along the
 field lines and reaches the stellar surface at the magnetic
 poles. The hatched regions represent shocks in front of and behind
 the subsonic propeller moving through the dense environment.}
          \label{F1}
   \end{figure}

Within this picture (see Fig.\,\ref{F1}) the envelope is
quasi-static: the characteristic time of the drift of material
through the envelope in the radial direction is much larger than
the characteristic time of the turbulent motions. The rate of the
drift is significantly smaller than $\dot{M}_{\rm c}$, i.e. the
rate by which the circumstellar material overflows the outer edge
of the envelope compressing the envelope plasma (for discussion
see Ikhsanov \cite{i03}, and references therein).

Under the condition $P_{\rm s} > P_{\rm cd}$, the centrifugal
barrier is not effective: the plasma, penetrating from the
envelope into the magnetic field of the star, is able to flow
along the magnetic field lines and to accrete onto the stellar
surface. Therefore, an accretion process onto the stellar surface
can be expected. However, the rate of this process is limited to
the rate of plasma penetration into the magnetic field of the
star.

As shown by Arons \& Lea (\cite{al76a}) and Elsner \& Lamb
(\cite{el76}), the interchange (e.g. Rayleigh-Taylor)
instabilities of the magnetospheric boundary under the condition
$T_{\rm pl} \sim T_{\rm ff}$ are suppressed (we would like to note
that the observed luminosities of AXPs are a factor of 5--10 lower
than the critical luminosity for which Compton cooling is
effective). In this situation, the process of plasma penetration
into the magnetosphere is governed by the Bohm diffusion and/or
the reconnection of the field lines at the magnetospheric
boundary. An evaluation of the corresponding rates of penetration
has recently been reported by Ikhsanov (\cite{i01a}, see Eqs.~22
and 24). Using these results, one finds that the rate of plasma
penetration into the magnetosphere due to the reconnection of the
field lines is high enough for the X-ray luminosity of subsonic
propellers to be of the order of that observed from AXPs (see
Eq.~(24) in Ikhsanov \cite{i01a})
 \be
 L_{\rm acc} = \dot{M}_{\rm rec} \frac{GM_{\rm ns}}{R_{\rm ns}}\ \sim\
 4 \times\ 10^{35}\ {\rm erg\,s^{-1}}\ \times
 \ee
 \bdm
\left[\frac{M_{\rm ns}}{1.4M_{\sun}}\right] \left[\frac{R_{\rm
ns}}{10^6\,{\rm cm}}\right]^{-1} \left[\frac{\alpha_{\rm
r}}{0.1}\right] \left[\frac{\lambda_{\rm m}}{0.1 R_{\rm m}}\right]
\left(\frac{\dot{M}_{\rm c}}{3\times 10^{17}\,{\rm
g\,s^{-1}}}\right).
 \edm
Here $\dot{M}_{\rm rec}$ is the rate of plasma penetration from
the envelope into the magnetosphere due to the reconnection
process, and $R_{\rm ns}$ is the radius of the neutron star.
$\alpha_{\rm r}$ is the efficiency of the reconnection process,
which is normalized according to the results of modeling of the
reconnection processes in solar flares and in the Earth's
magnetopause (see Priest \& Forbes \cite{pf00} and references
therein). $\lambda_{\rm m}$ is the average scale of plasma
inhomogeneities at the base of the envelope, which is normalized
following Arons \& Lea (\cite{al76b}), and Wang \& Robertson
(\cite{wr85}).

This indicates that the radiation of subsonic propellers situated
in a strong wind is powered mainly by the reconnection-driven
accretion of material onto the stellar surface. At the same time,
from 5 to 10\,percent of the radiation emitted by these objects
are expected to be contributed by the hot envelope surrounding the
magnetosphere (see Eq.~\ref{lssp}). Therefore, within the
considered approach the existence of at least two spatially
separated sources of the X-ray emission can be expected.

A detailed analysis of the radiation spectrum of subsonic
propellers is beyond the scope of this paper and will be presented
in a forthcoming paper.

   \section{Age}

The age of subsonic propellers can be expressed as the
superposition of pulsar-like and supersonic propeller spin-down
time scales.

  \subsection{Pulsar-like spin-down time scale}

The period at which the neutron star switches its state from
ejector to supersonic propeller is (see Eq.~6 in Ikhsanov
\cite{i01a})
   \be\label{pmd}
P_{\rm md} = 0.2\ \mu_{30.3}^{1/2}\ \dot{M}_{17.5}^{-1/4}\
V_7^{-1/4}\ {\rm s}.
     \ee
The spin-down rate of the star in the ejector state is
 \be
 \dot{P} = \frac{8 \pi^2 \mu^2}{3 c^3 I P}.
 \ee
Combining these equations, one finds the spin-down time scale
($\tau_{\rm a}=P/2\dot{P}$) as
 \be
 \tau_{\rm a} \simeq 1.2 \times 10^5\ \mu_{30.3}^{-1}\ I_{45}\
 \dot{M}_{17.5}^{-1/2}\ V_7^{-1/2}\ {\rm yr},
 \ee
where $V_7=V_{\rm rel}/10^7\,{\rm cm\,s^{-1}}$.

   \subsection{Supersonic propeller spin-down time scale}

The spin-down rate of a star in the state of supersonic propeller
is
 \be\label{dotpsups}
 \dot{P} = \frac{P^3 L_{\rm sups}}{4 \pi^2 I},
 \ee
where $L_{\rm sups}$ is the rate of the rotational energy loss by
the star in the state of supersonic propeller. In the general
case, this value can be limited to $\la \dot{M}GM/R_{\rm m}$.
Therefore, combining Eqs.~(\ref{pmd}) and (\ref{dotpsups}), one
can express the lower limit to the spin-down time scale of stars
in the supersonic propeller state as
 \be
 \tau_{c} \simeq 8 \times 10^4\ \mu_{30.3}^{-3/7}\ I_{45}\
 \dot{M}_{17.5}^{-11/14}\ V_7^{1/2}\ {\rm yr}.
 \ee
Although our estimate represents the absolute minimum for
$\tau_{\rm c}$, the real value of this parameter is unlikely to be
significantly larger. In particular, as shown by Lipunov \& Popov
(\cite{lp95}), the duration of the supersonic propeller state
under all reasonable conditions turns out to be smaller than the
spin-down time scale of the star in the state of ejector. On this
basis, one can limit the age of the star as it has evolved to the
state of subsonic propeller as
 \be
(\tau_{\rm a} + \tau_{\rm c}) \la \tau \la 2\tau_{\rm a}
 \ee

   \subsection{Subsonic propeller spin-down time scale}

Finally, the spin-down time scale of the star in the subsonic
propeller state can be expressed following  Ikhsanov (\cite{i01b},
Eqs.~10), as
   \be\label{taud}
\tau_{\rm d} \simeq 2\times 10^3\ \mu_{30.3}^{-2}\ I_{45}\ P_{5}\
m_{1.4}\ {\rm yr},
  \ee
where $m_{1.4} = 1.4\,M_{\sun}$.

$\tau_{\rm d}$ is small compared to the value of both, $\tau_{\rm
a}$ and $\tau_{\rm c}$. Hence, the age of the neutron stars with
the parameters derived in Sect.\,2 is determined mainly by the
pulsar-like spin-down time scale and can be limited to $<
2\tau_{\rm a}$. In order to fit the currently adopted evaluations
of the ages of AXPs, one has to suggest that in a previous epoch
the conditions in the circumstellar environment were slightly
different. In particular, one could assume that the strength of
the wind in the previous epoch was about $10^{18}\,{\rm
g\,s^{-1}}$, which corresponds to the accretion process at the
Eddington limit. On the other hand, one can also envisage a
situation in which the value of the relative velocity $V_{\rm
rel}$ is slightly larger than 100\,${\rm km\,s^{-1}}$ (this
normalization of the velocity comes from the condition for the
self-consistency of the supersonic propeller model, see Eq.~7 in
Ikhsanov \cite{i02}). In this case the age of subsonic propellers
turns out to be close to the typical ages of the SNRs with which
they are presently associated.

  \section{Discussion}

As shown above, some basic features of AXPs can be explained in
terms of neutron stars in the state of subsonic propeller provided
their mass and magnetic moment are $M\sim 1.4\,M_{\sun}$ and $\mu
\sim 2\times 10^{30}\,{\rm G\,cm^3}$, and the strength of the wind
is $\dot{M}_{\rm c} \sim 3\times 10^{17}\,{\rm g\,s^{-1}}$. The
required values of mass and dipole magnetic moment are typical for
neutron stars. The required strength of the wind, however, is
rather high and this condition appears to be a strong restriction
on the applicability of the proposed scenario. Indeed, for the
wind strength $\sim 10^{17}\,{\rm g\,s^{-1}}$ to realize, the
neutron star should be embedded in a medium of density
 \be
 \rho \sim 10^{-16}\ \dot{M}_{17}\ M_{1.5}^{-2}\ V_7^3\ {\rm
 g\,cm^{-3}}.
 \ee
Even in the case of a low relative velocity $V_{\rm rel} \sim
10^6\,{\rm cm\,s^{-1}}$ (which is about the sound speed at the
accretion radius of the star), the required number density of the
surrounding gas, $N_{\rm e} \sim 10^5\,{\rm cm^{-3}}$, remains
significantly larger than the average density of material in SNRs
and ISM. Hence, for the proposed scenario to be applicable one has
to assume that AXPs are embedded in relatively dense regions.

What are the physical conditions in these regions\,? As follows
from X-ray observations, the column density in the observed AXPs
is $(0.4-2)\times 10^{22}\,{\rm cm^{-2}}$ (see e.g. Table\,2 in
Mereghetti et~al. \cite{mcis02}). This indicates that the column
density in the dense region is limited to
 \be\label{coldens}
 N_{\rm e} L \la 4 \times 10^{21}~{\rm cm^{-2}},
 \ee
where $N_{\rm e}$ is the number density of the material in the
dense region and $L$ is the size of the region along the line of
sight.

If we require the life time of the accreting X-ray source to
exceed the spin-down time scale of the subsonic propeller, one
gets the following condition:
 \be\label{mass-tot}
 N_{\rm e} m_{\rm p} \Phi \ga \frac{L_{\rm acc} R_{\rm
 ns}}{GM_{\rm ns}}\ \tau_{\rm d},
 \ee
where $\Phi$ is the volume of the dense region.

Finally, the total flux emitted by the material accumulated in the
dense region should not exceed the fluxes detected from the
infrared counterparts of AXPs. Under the conditions of interest,
the dense regions are optically thin at the frequencies
$10^{14}-10^{15}$\,Hz. Hence, using the data presented in Fig.\,2
of Israel et~al. (\cite{isc04}) one finds:
 \be\label{flux}
 N_{\rm e}^2 T^{1/2} \Phi \la 6 \times 10^{58},
 \ee
where $T$ is the average temperature of the material in the dense
region.

Combining Eq.~(\ref{coldens}) -- (\ref{flux}) and setting $T \sim
2000$\,K (the error of a factor of 2 in this parameter does not
affect the final conclusions) we find that the subsonic propeller
approach to the interpretation of AXPs is applicable if the
corresponding X-ray sources are embedded in regions of volume
$\Phi \sim 5\times 10^{42}\,{\rm cm^3}$, and average number
density $N_{\rm e} \sim 10^7\,{\rm cm^{-3}}$. The size of the
regions along the line of sight is limited to $L \la 7\times
10^{13}$\,cm. Under these conditions the appearance of subsonic
propellers is in agreement with properties of AXPs in both the
X-rays and the infrared part of the spectrum.

The question about the possible origin of such regions is rather
complicated. In contrast to models built around the assumption
that neutron stars in AXPs are in the state of accretor, the
gravitational potential of the star within our scenario does not
play a central role in the formation of its dense environment.
Indeed, the required size of the environment significantly exceeds
the accretion radius of the star. This indicates that the
formation of the dense region is rather independent of the
properties of the neutron star itself, but is related to the
parameters of the supernova explosion or the properties of its
progenitor or both.

Three possible scenarios for the formation of dense environments
around AXPs has recently been outlined by Marsden et~al.
(\cite{mlrh01}): ``fall-back'' disk accretion, ``pushed-back''
disk accretion, and accretion involving high-velocity neutron
stars. The fall-back disk accretion scenario within the subsonic
propeller approach can be rejected since the material stored in
the disk is assumed to be gravitationally bound to the star (this
disk is situated within the accretion radius of the star, see e.g.
Chatterjee et~al. \cite{chn00} and references therein).

The scenario in which a high-velocity neutron star captures
material from the co-moving ejecta can be rejected on the same
basis. Indeed, the number density in the expanding ejecta is $n
\propto t^{-3}$ and therefore the life-times of AXPs within the
subsonic propeller scenario turn out to be too short. The
life-time could be longer if the material captured by the star
were stored in a disk within its gravitational radius, but this
assumption contradicts the basic statements of our model.

Within the pushed-back disk scenario the dense environment of the
star is expected to form on a time scale of $10^3$\,yr. The total
amount of material stored in such a disk is about $0.4\,M_{\sun}$
(see Marsden et~al. \cite{mlrh01}, and references therein) and the
disk is not gravitationally bound to the neutron star.
Unfortunately, the basic properties of such a disk have been
investigated rather poorly. For this reason, a detailed analysis
of the applicability of this model to the subsonic propeller
approach is presently not possible. Nevertheless, the above
mentioned estimates of the mass and the time scale indicate that
the pushed-back disk scenario could be a solution for the problem
of the origin of the dense environment required in our model.

It should also be noted that both the identification of the
infrared counterparts of AXPs (Israel et~al. \cite{isc04}) and the
conclusion about the compact geometry of their host SNRs (Marsden
et~al. \cite{mlrh01}) give some hints that there exists a dense
environment surrounding these objects. In the light of this, the
requirements of our model do not seem to be artificial but turn
out to be rather reasonable from the observational point of view.

   \section{Conclusion}

The appearance of magnetized ($B\sim 3\times 10^{12}$\,G) neutron
stars in the state of subsonic propeller, which are situated in a
strong ($\dot{M}_{\rm c} \sim 3 \times 10^{17}\,{\rm g\,s^{-1}}$)
wind, is similar to that of AXPs in several important aspects.
Namely, the rotational periods of these objects are limited to the
range of 5--15\,s, and the spin-down rates are about $7 \times
10^{-11}\,{\rm s\,s^{-1}}$. The emission of stars in this state is
partly (5--10\%) contributed to by the hot envelope surrounding
its magnetosphere and mainly (90--95\%) by the process of
accretion of material onto the stellar surface. The efficiency of
the accretion process, which is governed by the reconnection of
the magnetic field lines at the magnetospheric boundary, is high
enough for the luminosity of the considered objects to be of order
a few\,$\times 10^{35}\,{\rm erg\,s^{-1}}$. Finally, the estimated
ages of neutron stars under the conditions of interest are close
to $10^5$\,yr.

The above derived parameters allow us to suggest that the
interpretation of AXPs within the subsonic propeller approach is
rather promising for the future investigations.

 \begin{acknowledgements}
 We would like to thank the anonymous referee for useful comments.
The work was partly supported by the Russian Foundation of Basic
Research under grant 03-02-17223a, and the State Scientific and
Technical Program ``Astronomy''.
\end{acknowledgements}


\begin{thebibliography}{}
 \bibitem[1976a]{al76a}
 Arons J., Lea S.M., 1976a, ApJ 207, 914
\bibitem[1976b]{al76b}
 Arons, J., Lea, S.M. 1976b, ApJ 210, 792
 \bibitem[1981]{dp81}
 Davies R.E., Pringle J.E., 1981, MNRAS 196, 209
\bibitem[2000]{chn00}
 Chatterjee, P., Hernquist, L., Narayan, R. 2000, ApJ, 534, 373
\bibitem[2000]{cgp00}
 Colpi, M., Geppert, U., Page, D. 2000, ApJ, 529, L29
\bibitem[2003]{ea03}
 Eksi, K.Y., Alpar, M.A. 2003, ApJ, 599, 450
 \bibitem[1976]{el76}
 Elsner R.F., Lamb F.K., 1976, Nature 262, 356
\bibitem[2001a]{i01a}
 Ikhsanov N.R., 2001a, A\&A, 375, 944
\bibitem[2001b]{i01b}
 Ikhsanov N.R., 2001b, A\&A, 368, L5
\bibitem[2002]{i02}
 Ikhsanov N.R., 2002, A\&A, 381, L61
\bibitem[2003]{i03}
 Ikhsanov N.R., 2003, A\&A, 399, 1147
\bibitem[2004]{isc04}
 Israel, G.L., Stella, L., Covino, S., et~al. 2004, in ``Young
   Neutron Stars and Their Environments'', eds. F.\,Camilo and B.M.\,Gaensler,
   IAU Symp., 218, in press (astro-ph/0310482)
\bibitem[1995]{lp95}
 Lipunov, V.M., Popov, S.B. 1995, Astron. Reports, 39, 632
\bibitem[2001]{mlrh01}
 Marsden, D., Lingenfelter, R.E., Rotschild, R.E., Higdon, J.C.
 2001, ApJ, 550, 397
\bibitem[2002]{mcis02}
 Mereghetti, S., Chiarlone, L., Israel, G.L., Stella, L. 2002, in
 ``Neutron Stars, Pulsars and Supernova Remnants'', eds.
 W.\,Becker, H.\,Lesch and J.\,Tru\"mper,
 MPE\,Report, v.\,278, p.\,29
\bibitem[2000]{pf00}
 Priest, E.R., Forbes, T.G. 2000, ``Magnetic reconnection: MHD
 theory and applications'', Cambridge University Press
\bibitem[2002]{pm02}
 Psaltis, D., Miller, M.C. 2002, ApJ, 578, 325
\bibitem[1996]{td96}
 Thompson, C., Duncan, R.C. 1996, ApJ, 473, 322
\bibitem[1985]{wr85}
 Wang, Y.-M., Robertson, J.A. 1985, A\&A 151, 361
 \end{thebibliography}
\end{document}